\documentclass[pra,aps,floatfix,twocolumn,showpacs,reprint,superscriptaddress,linenumbers]{revtex4}
\usepackage{lipsum}
\usepackage{subfigure}
\usepackage{hyperref} 
% ...
\usepackage[T1]{fontenc}             
\usepackage[applemac]{inputenc} 
\usepackage{mathptmx}                
\usepackage[scaled=.90]{helvet}
\usepackage{graphicx}                
\usepackage{color}                
\usepackage{setspace}                
\usepackage[sort&compress]{natbib}   
\usepackage[loose,nice]{units}       
\usepackage{tabularx}                
\usepackage{booktabs}                
\usepackage{amsmath}
\usepackage{ulem}
\usepackage{amssymb}
\usepackage{url}
\newcolumntype{C}{>{\centering\arraybackslash}X}
\newcolumntype{L}{>{\raggedleft\arraybackslash}X}
\newcolumntype{R}{>{\raggedright\arraybackslash}X}

\usepackage{graphicx}
\usepackage{textgreek}
\usepackage{sidecap}
\usepackage{floatrow}
\usepackage{sidecap}  
\usepackage{dcolumn}
\usepackage{bm}
\usepackage{hyperref}%
\hypersetup{
    colorlinks=true,
    linkcolor=blue,
    citecolor=blue,
    filecolor=magenta,      
    urlcolor=blue,
}

\begin{document}


\date{\today}

\title[]{Infrared single-cycle pulse induced high-energy plateaus in high-order harmonic spectroscopy}

\author{Abdelmalek Taoutioui}
\affiliation{Institute for Nuclear Research, ATOMKI, Debrecen, Hungary\\}
\affiliation{Physique du Rayonnement et des Interactions
Laser-Mati\`ere, Facult\'e des Sciences. Universit\'e Moulay Ismail,
B.P. 11201, Zitoune, Meknes, Morocco\\}
\author{Hicham Agueny}
\email{hicham.agueny@uib.no}
\affiliation{Department of Physics and Technology, Allegt. 55,
University of Bergen, N-5007 Bergen, Norway}

\begin{abstract}
Motivated by the emerging experiments [e.g. \textit{Z. Nie et al. Nat. Photon. \textbf{12}, 489 (2018)}] on producing infrared (IR) single cycle pulses in the spectral region 5 - 14 $\mu m$, we theoretically investigate their role for controlling high-order harmonic generation (HHG) process induced by an intense near-infrared (NIR) multi-cycle pulse ($\lambda$ = 1.27 $\mu m$). The scenario is demonstrated for a prototype of the hydrogen atom by numerical simulations of the time-dependent Schr\"odinger equation. In particular, we show that the combined pulses allow one to generate even-order harmonics and most importantly to produce high-energy plateaus and that the harmonic cutoff is extended by a factor of 3 compared to the case with the NIR pulse alone. The emerged high-energy plateaus is understood as a result of a vast momentum transfer from the single-cycle field to the ionized electrons while travelling in the NIR field, and thus leading to high-momentum electron recollisions. We also identify the role of the IR single-cycle field for controlling the directionality of the emitted electrons via the IR-field induced electron displacement effect. We further show that the emerged plateaus can be controlled by varying the relative carrier-envelope phase between the two pulses as well as their wavelengths. Thus, our findings open up new perspectives for time-resolved electron diffraction using an IR single-cycle field-assisted high-harmonic spectroscopy. 
\end{abstract}


\maketitle


\section{INTRODUCTION}
High harmonic generation (HHG) is a coherent process that plays a key role in producing ultrashort coherent light in the extreme ultraviolet (XUV) and soft x-ray range~\cite{Popmintchev2010, Popmintchev2018}, generating attosecond pulses~\cite{Corkum2007,Krausz2009}, and now has been growing rapidly for imaging molecular orbitals~\cite{Itatani2004, Lin2010,Peng2019} and for temporal characterization of ultrafast processes~\cite{Leeuwenburgh2013,Silva2019}.

The underlying physics of HHG is well understood on the basis of the three-step model~\cite{Corkum1993}, in which tunnelling, acceleration and recombination of the electrons are the fundamental steps responsible of the generation of high-order harmonics. Thus, enhancing the process relies on controlling these basic steps, and its efficiency manifests by an extension of the harmonic cutoff and/or by an increase of the harmonic yield~\cite{Lara2016}. Here, extensive theoretical works evoking different schemes have been devoted to achieve this goal. For instance, the use of chirped laser pulses~\cite{Carrera2007,Lara2016,Peng2018}, and the spatial inhomogeneity of the laser field~\cite{Kim2008h, Ciappina2012} have been shown to lead to an extension of the cutoff region and or enhancing the intensity of HHG. Other schemes involving static electric fields have been investigated for controlling HHG~\cite{Bao1996,Wang1999,Borca2000,Odzak2005}. Here although even harmonics can be generated~\cite{Bao1996}, the extension of the harmonic cutoff and the enhancement of the yield is very limited by the the strength of the static field, which in principle cannot be too high in an experiment. In addition, the fields are not appropriate for a precision control of the electron motion in time-domain. Alternative schemes to this issue is the use of color mixing, in which the field-induced HHG is assisted by weaker pulses~\cite{Schiessl2006,Peng2017}. This includes the assisting XUV~\cite{Fleischer2008a, Fleischer2008b, Sarantseva2018,Sarantseva2020}, UV~\cite{Popruzhenko2010}, VUV~\cite{Miller2014} and THz pulses~\cite{Kovacs2012}. The origin of the enhancement in these color-mixing schemes was different depending of the assisting field. For instance, in the presence of a weak XUV pulse, the enhancement was linked either to the XUV-induced ac-Stark effect in the ground state~\cite{Fleischer2008b} or to the absorption of XUV photons during the recombination step~\cite{Sarantseva2018}. On the other hand, the enhancement in presence of a THz field was found to be caused by the modulation of the accumulated dipole phase, which results in constructive interferences of the dipole emissions (i.e. phase matching)~\cite{Kovacs2012}.

Despite the extensive works in the field of HHG, there is still major needs for seeking new schemes capable of pushing the harmonic cutoff to higher energies with or without generating high-energy plateaus. It is thus the purpose of the present work to reveal a so far unexplored route for enhancing and controlling the HHG process, thus complementing the existing schemes and adding new insights to the general field of strong-field and attosecond physics. The proposed scheme is based on introducing a weak infrared (IR) single-cycle pulse combined with an intense laser pulse. The characteristic feature of the single-cycle pulse relies on a high-momentum transfer to electrons, leading to their displacement mainly in a single-direction. This has been discussed in the context of a THz single-cycle field interacting with Rydberg atoms~\cite{Jones2014,Robicheaux2014} (see also~\cite{Robicheaux2015,Agueny2016,Misha2017a}) and its underlying physics has been shown to be valid in the ultrafast regime~\cite{Misha2017b,Agueny2020a}, in which a coherent displacement of the electron wavepacket was demonstrated~\cite{Agueny2020a}.       
 
In this context, there has been recently a significant progress in developing schemes capable of producing single-cycle pulses in the infrared spectral range~\cite{Krauss2010,Balciunas2015,Nie2018} for controlling ultrafast phenomena in gases and solids~\cite{Nomura2012,Krogen2017,Liang2017,Nie2018,Chen2019,Thiele2019,Zhu2019,Hwang2019,Zhu2020}. This specific interest is motivated by the properties of these pulses in isolating the electron motion in strong-light matter interaction~\cite{Nie2018}. Some application examples include the generation of high-energy electrons~\cite{Li2016} and electron currents in the petahertz regime~\cite{Rybka2016}, sub-femtosecond control of the nonlinear response of bound electrons in atoms~\cite{Hassan2016}, precise control of the electron transport in plasmonic gaps~\cite{Ludwig2020}, and very recently the sub-femtosecond control of freely propagating electron beam~\cite{Morimoto2020} was demonstrated experimentally. 

It is thus timely to guide these experimental efforts and provide new insights into the role of IR single-cycle pulses in controlling coherent processes such as HHG. We aim in this work at studying to which extent the presence of a weak IR single-cycle pulse modifies the HHG process. Here, we use an intense near-infrared (NIR) multi-cycle pulse to generate high-order harmonic components and by exploiting the characteristic features of the combined pulses we show that high-energy plateaus can emerge. The emergence of this high-energy phenomenon is itself an important aspect towards establishing a new spectroscope for time-resolved electron diffraction. Specifically, we find that the characteristics of the plateau manifest by an extension of the harmonic cutoff by a factor of 3 compared to the case with only the NIR pulse. The origin of the phenomenon is linked to a displacement of the electrons caused by the single-cycle field, which in turn get further accelerated to higher momenta following a unidirectional path, thus leading to high-energy electron recollisions. We also identify the role of the field for generating even-order harmonics and controlling the directionality of the ionized electrons in the forward-backward direction. Furthermore, we show that varying the relative optical phase between the two pulses as well as the wavelength of the IR single-cycle pulse modifies dramatically the high-harmonic spectrum, and consequently, the extension of the  plateaus can be controlled. Our study is based on numerical simulations of the time-dependent Schr{\"{o}}dinger equation (TDSE). Although, the calculations are based on a one-dimensional (1D) model, the basic physics involved during the electron dynamics is verified using a three-dimensional (3D) model.

This paper is organized as follows. In Sec. \ref{theory} we present our theoretical models based on 1D- and 3D-TDSE, including a short description of our numerical methods for solving the TDSE and for calculating the HHG spectrum. In Sec. \ref{results} we present our results for HHG spectra produced by the combined pulses, and outline the effect of the presence of the single-cycle field for controlling the HHG process, and discuss the physics behind the emerged effects. The findings are supported by an analysis based on the time-evolution of the ionized electrons presented in momentum space and by the Gabor time-frequency analysis. We also discuss how the relative optical phase and the single-cycle wavelength affect the HHG spectrum. Finally, in Sec. \ref{conclusions} we summarize our results on the impact of using the assisting single-cycle pulse on HHG. Atomic units (a.u.) are used throughout this paper unless otherwise specified.

\section{THEORETICAL MODELS}\label{theory}

The TDSE governing the electron dynamics induced by coherent light pulses is written as 
\begin{equation}\label{tdse}
\Big[ H_0 + H_I(t) - i\frac{\partial}{\partial t}\big]|\psi(t) \rangle=0,
\end{equation}
where $H_0=-\frac{\nabla^{2}}{2} +V(r)$ is the field-free Hamiltonian with the potential interaction $V(r)$. The time-dependent interaction $H_I(t)$ is treated in the length gauge and is described within the dipole approximation. Without loss of generality, the electric fields are considered to be linearly polarized along the $z$-direction. Here, we consider a two-color scheme, in which the interaction $H_I(t)$ can be expressed as
\begin{equation}\label{tdi}
H_I(t) = -\mathbf{z} \cdot (\mathbf{F_{NIR}}(t) + \mathbf{F_{IR}}(t)),
\end{equation}
where $F_{NIR}(t)$ and $F_{IR}(t)$ describe, respectively, the NIR multi-cycle pulses and IR single-cycle pulse. In our calculations, we use the following form for the NIR pulse
\begin{equation}\label{F1}
F_{NIR}(t) = E_{NIR}\cos^2(\pi t/\tau_{NIR})\cos(\omega_{NIR} t + \delta\phi),
\end{equation}
while the IR single-cycle pulse is expressed as
\begin{equation}\label{F2}
F_{IR}(t) = E_{IR}\mathrm{e}^{-t^{2}/(2\sigma_{IR}^{2})}15.53\frac{t}{\tau_{IR}},
\end{equation}
where $\sigma_{IR}=\tau_{IR}/4\sqrt{2\ln(2)}$ is the width of the Gaussian function in Eq. (\ref{F2}). Here, $\tau_{IR}=2\pi/\omega_{IR}$, and $\tau_{NIR}=2\pi T_c/\omega_{NIR}$ are the total duration of the IR and NIR pulses. $T_c$ is the total number of cycles of the NIR pulse. The parameters $\omega_{NIR}$ ($\omega_{IR}$) and $E_{NIR}$ ($E_{IR}$) are respectively, the central frequency and the amplitude of the NIR pulse (IR pulse). The amplitude $E_{i}$ $(i=NIR,IR)$ is related to the peak intensity via the relation $I_i=E_i^{2}$. The pulses in Eqs. (\ref{F1}) and (\ref{F1}) satisfy the condition $\int_{t_i}^{t_f} F_i(t) dt $ = 0, where $\tau_{i}=t_f-t_i$, and their form is depicted in Fig. \ref{fig1}(a). 

We calculate the HHG spectrum $H(\omega)$ by carrying out the Fourier transform of the expectation value of the dipole acceleration along the $z$-axis
\begin{equation}\label{H}
H(\omega) = |D_z(\omega)|^2,
\end{equation} 
where $D_z(\omega)$ is defined by
\begin{equation}\label{Dzw}
D_z(\omega) = \frac{1}{\sqrt{2\pi}} \int_{-\infty}^{+\infty} <D_z(t)> \mathrm{e}^{-i\omega t} dt,
\end{equation} 
and the time-dependent expectation value of the dipole acceleration $<D_z(t)>$ is written as~\cite{bandrauk2009}
\begin{equation}\label{Dzt}
D_z(t) = \langle \psi(t)| \frac{\partial V(r)}{\partial z} |\psi(t)\rangle + F(t),
\end{equation}  
where $F(t)$ is the combined NIR and IR pulses. We finally define the HHG power spectrum by
\begin{equation}\label{Pw}
P(\omega) = \frac{H(\omega)}{\tau_{NIR}^2 \omega_{NIR}^4}, 
\end{equation}  
in which a window function of a Gaussian form $\exp{[-(t-t_0)^2/(2\sigma^2)]}$ centred at $t_0$ and having the width $\sigma$=5.77/$\omega_{NIR}$ is used. Note that convolution with a Gaussian window allows in general a faster decaying of a desired function, which in our case is the dipole accelerator [cf. Eq. (\ref{Dzt})], at the boundaries. Its utilization prior to a Fourier transform analysis is a well-know technique that spans all physical sciences as it is useful to highlight tiny effects and has been used extensively in the context of HHG.

For further analysis we calculate the laser-induced electron current and the expectation value of the energy. These are given, respectively by 
\begin{equation}\label{current}
j(t) = -\Re[\langle \psi(t)| \hat{p}_z |\psi(t)\rangle],
\end{equation}  
and 
\begin{equation}\label{Energy}
<E>(t) = -0.5\langle \psi(t)| \frac{\partial^2}{\partial z^2} |\psi(t)\rangle.
\end{equation}  

The ionization wavefunction is also calculated using the same methodology as described in \cite{Agueny2018,Agueny2020b}
\begin{equation}\label{ionization}
|\psi_{ioniz}(t)\rangle = |\psi(t)\rangle - \sum_i |\phi_i\rangle \langle\phi_i|\psi(t)\rangle.
\end{equation}
Here, the sum in Eq. (\ref{ionization}) covers the important bound states $|\phi_i\rangle$. We have checked that the extraction of the first 10 bound states is enough for the convergence of the ionization wavefunction. 

For solving the TDSE Eq. (\ref{tdse}), we use both 1D- and 3D-models. And because of the extensive calculations involving the field range parameters, and which are performed on large spatial and temporal grids, we apt for a 1D-model. We have however verified the validity of our results by performing calculations based on a 3D-model. Therefore, the physical mechanisms discussed here hold for a realistic scenario. We stress that extensive theoretical works have been carried out using a 1D-model, and which was shown to capture the basic physics involved in an experiment. 

In our numerical simulations, we consider a prototype of the hydrogen atom initially prepared in the ground state. This initial state is obtained by propagating in the imaginary time. The time evolution of the electronic wave function $|\psi(t)\rangle$, which satisfies the TDSE [cf. Eq. (\ref{tdse})], is solved numerically using a split-operator method. In the case of a 1D-model, in which a soft potential of the form $V(z)=-1/\sqrt{z^2 + 2.0}$ is used, the method is combined with a fast Fourier transform (FFT) algorithm. For 3D calculations, the atomic potential has the form $V(z,\rho)=-1/\sqrt{z^2 + \rho^2}$ and the TDSE is solved in cylindrical coordinates by combining the FFT along the $z$-axis, as in the 1D-model, and the stable Cayley transform with use of the three-point finite difference for the discretisation of the kinetic energy operator in the $\rho$ coordinates (further details about the numerical method can be found in~\cite{Agueny2018}). 

The parameters of the spatial grid along the $z$-direction are kept the same for both the 1D- and 3D-model. Here, calculations are carried out in a grid of size $L_z$ = 8192 a.u. and $L_{\rho}$ = 1230 a.u., respectively, along the $z$- and $\rho$-axes, with the spacing grid $dz$ = 0.25 a.u. and $d\rho$ = 0.15 a.u., i.e. $n_z$=32768 and $n_{\rho}$= 8192 grid points. The time step used in the simulation is $\delta t$=0.02 a.u.. The convergence is checked by performing additional calculations with twice the size of the box and a smaller time step. An absorbing boundary is employed to avoid artificial reflections, but without perturbing the inner part of the wave function. The boundary is chosen to span 10\% of the grid size in each direction.

\section{RESULTS AND DISCUSSION}\label{results}

\begin{figure}[h]
\centering
\includegraphics[width=8cm,height=6.5cm]{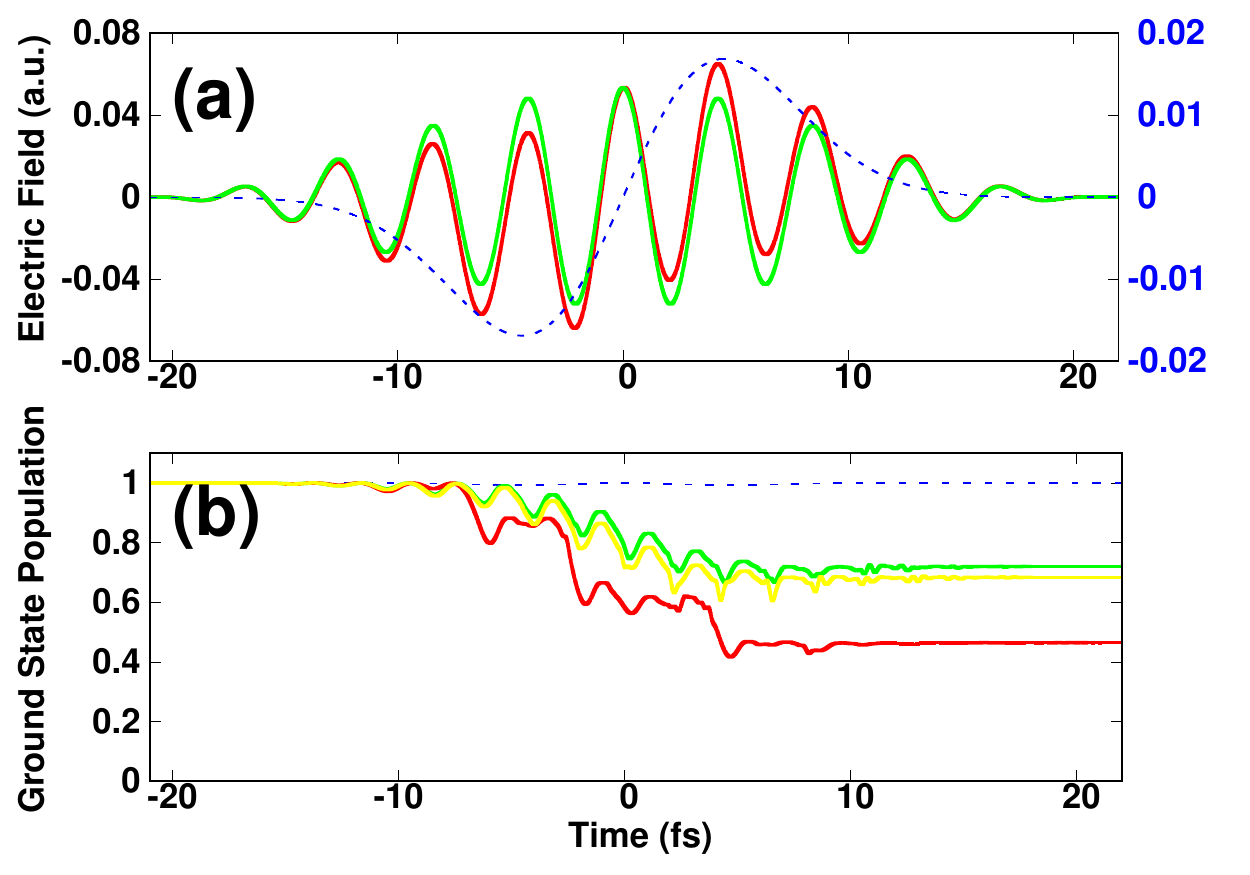}
\caption{\label{fig1} (Color online). (a) Laser pulses. (b) 1D-calculations of the temporal evolution of the population of the ground state induced by: the NIR pulse alone (green curve), the IR pulse alone (blue dashed line), and by the combined pulses (red curve). The corresponding pulses are shown in the top with the same colors. The parameters of the NIR pulse are: $\lambda_{NIR}=$ 1.27 $\mu m$, $T_c=$ 10 cycles, $\delta\phi=$ 0 and $I_{NIR}=$ 1$\times$10$^{14}$ W/cm$^2$. The parameters of the IR pulse are: $\lambda_{IR}=$ 12.7 $\mu m$ and $I_{IR}=$ 1$\times$10$^{13}$ W/cm$^2$. It is also shown, for reference, the population induced by only the NIR pulse, but for a peak intensity of 1.1$\times$10$^{14}$ W/cm$^2$ (yellow color).}
\end{figure}

We consider a two-color mixing scheme implemented numerically in order to control the HHG process. The scheme consists of an intense NIR multi-cycle pulse, which generates high-order harmonics and a weak IR single-cycle pulse introduced as an assisting field. The NIR pulse has 1.27 $\mu m$ central wavelength, 42.4 fs pulse duration and 1$\times$10$^{14}$ W/cm$^2$ peak intensity. The IR pulse has the same pulse duration but with a central wavelength of 12.70 $\mu m$, and the peak intensity is in the range [1$\times$10$^{10}$,1$\times$10$^{13}$] W/cm$^2$. Here, the choice of these intensities is such that the single-cycle field is introduced as a control tool and does not contribute to the ionization, but rather acts on the freely propagating electrons. This can be seen in Fig. \ref{fig1}(b) (blue dashed-line), in which the temporal evolution of the ground state population is almost unchanged. At the end of the pulse, the population is 0.994, and in the case of the NIR pulse alone is 0.72, while when combining both pulses the occupation becomes 0.46. We also present, for reference, the population in the case the IR single-cycle pulse is replaced by a weak NIR multi-cycle pulse having the same form and the wavelength as the intense one (i.e. the total peak intensity of the NIR is 1.1$\times$10$^{14}$ W/cm$^2$). The population is found to be 0.68: it shows a small depletion compared to the case when introducing the IR pulse (i.e. the induced population 0.46). Taking into consideration the characteristic feature of the single-cycle field; these preliminary results suggest that the fast depletion of the ground state might be caused by high-energy recollision of electrons that acquire high momentum from the single-cycle field. We elaborate this discussion in the following in connection with the HHG process. 

\begin{figure*}[ht]
\centering
\includegraphics[width=8cm,height=5cm]{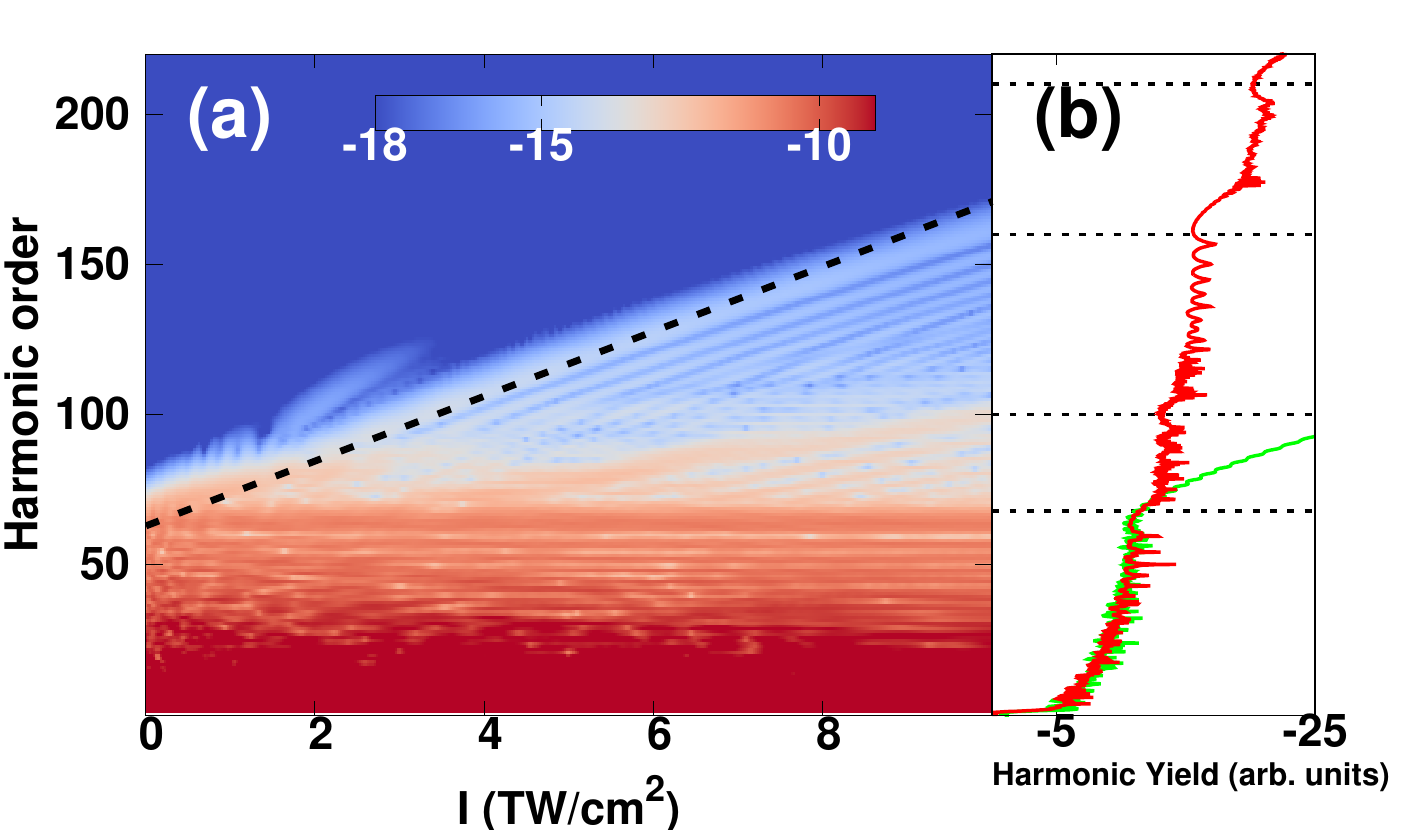}
\includegraphics[width=8cm,height=5cm]{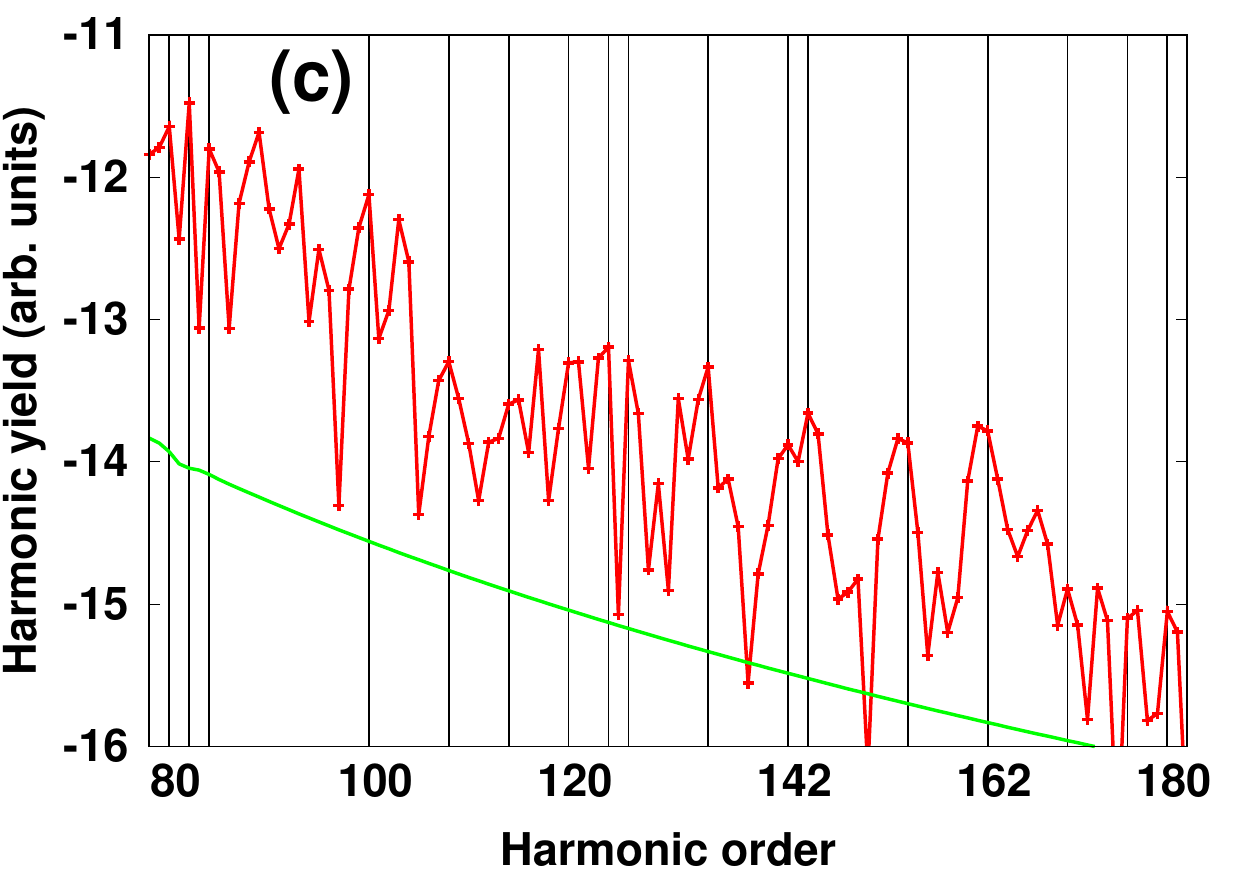}
\includegraphics[width=8cm,height=5cm]{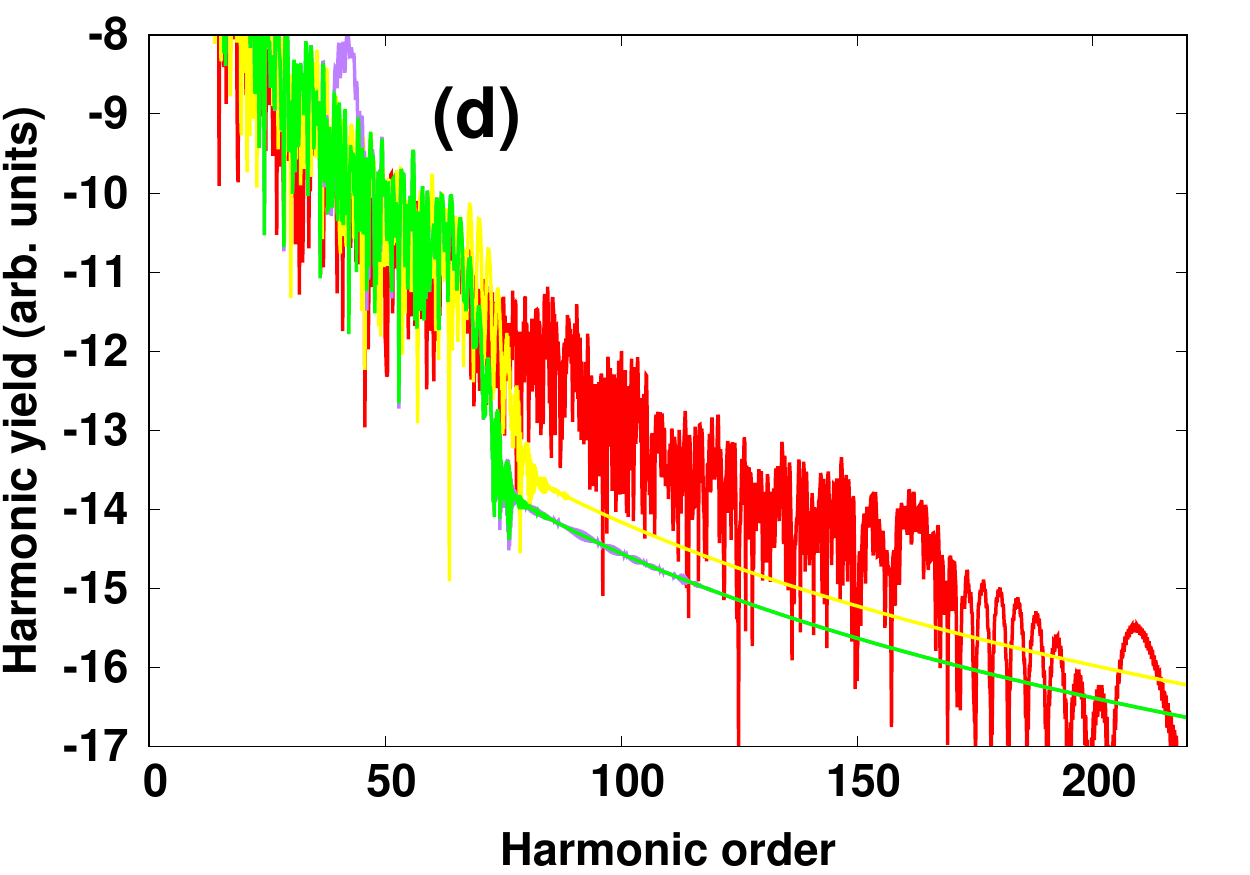}
\includegraphics[width=8cm,height=5cm]{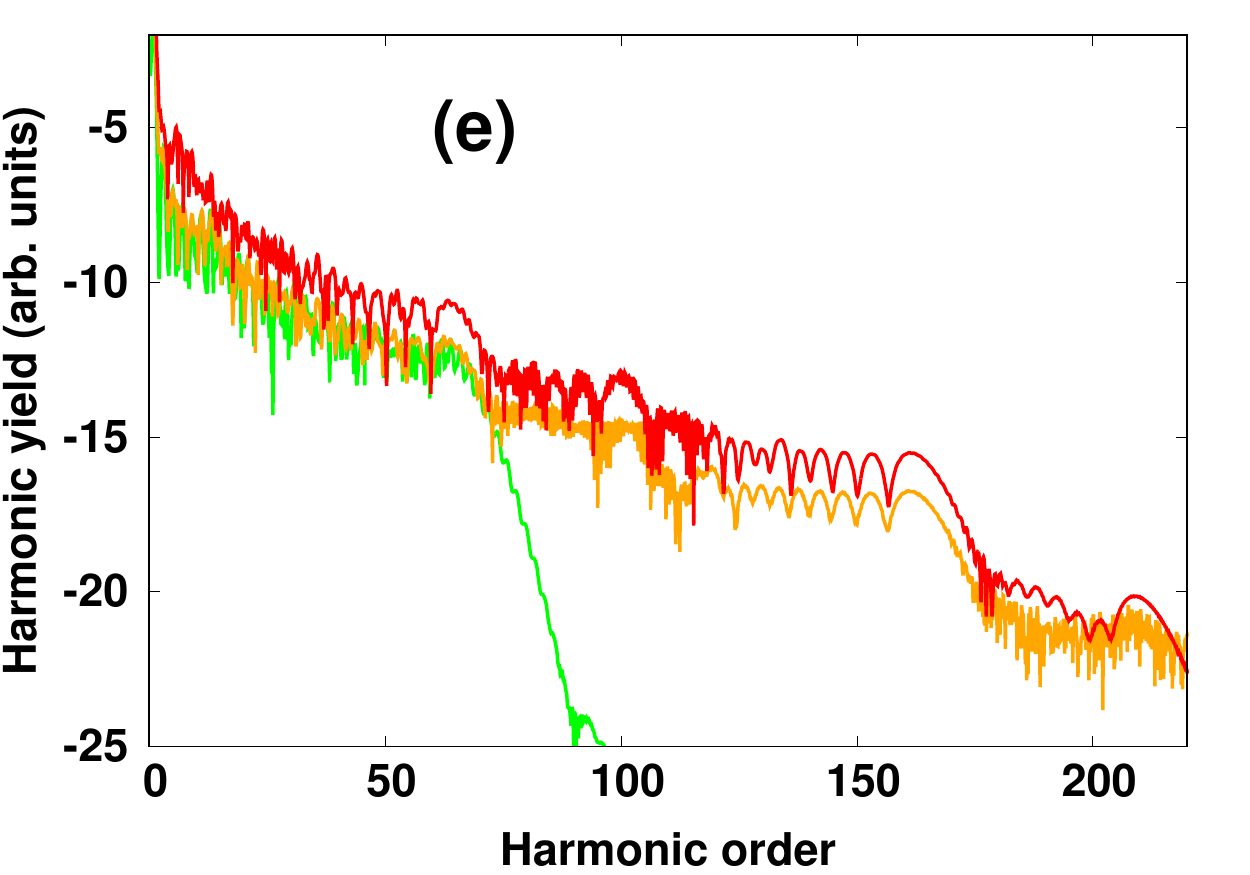}
\caption{\label{fig2} (Color online). (a)-(d) 1D-calculations of HHG. (a) Harmonic spectra as functions of the peak intensity of the IR single-cycle pulse. (b) Harmonic spectra obtained with the assisting IR pulse (red curve) and without (green curve). (c) Harmonic spectrum as in (b) but in the harmonic order range [80:180], in which even-order harmonics are indicated by vertical lines. (d) Harmonic spectra obtained with different assisting fields having the same peak intensity and without using a Gaussian window: (purple curve) XUV-assisting field ($\omega_{XUV}=$41 eV and $\tau_{XUV}=$1 fs), (yellow curve) NIR-assisting field ($\omega_{NIR}=$0.976 eV and $\tau_{NIR}=$ 42.4 fs ) and (red curve) IR single-cycle assisting field ($\omega_{IR}=$0.0976 eV, $\tau_{IR}=$42.4 fs). For reference, the spectrum obtained in the absence of the assisting field is also plotted (green curve). (e) 3D-calculations of the harmonic spectra as in (b): with the assisting IR pulse (orange curve) and without (green curve). For comparison, the spectrum obtained using the 1D-model is also shown (red curve). Yields are shown in log scale. The parameters of the NIR pulse are: $\lambda_{NIR}=$ 1.27 $\mu m$, $T_c=$ 10 cycles, $\delta\phi=$ 0 and $I_{NIR}=$ 1$\times$10$^{14}$ W/cm$^2$. The parameters of the IR pulse are: $\lambda_{IR}=$ 12.7 $\mu m$ and $I_{IR}=$ 1$\times$10$^{13}$ W/cm$^2$.}
\end{figure*}

In Fig. \ref{fig2}(a) we present the calculated HHG spectrum using the two-color scheme at different peak intensities of the IR single-cycle pulse. The calculations are based on a 1D-model. Here, the relative optical phase between the two pulses is fixed at 0. It is seen that the presence of the single-cycle pulse modifies dramatically the HHG spectrum. In particular, it can be seen that the cutoff of the harmonic order $N_c$ increases monotonically with increasing the peak intensity of the assisting IR single-cycle pulse. It extends by almost a factor of 3: It goes from $N_c=$60 in the absence of the IR pulse to $N_c=$170 when the assisting pulse is introduced. The cutoff scales linearly with the intensity $I_{IR}$, as indicated by the dashed-line in Fig. \ref{fig2}(a) and is found to follow the approximative formula 
\begin{equation}\label{Nc}
N_c= [3.17\frac{\tilde{I}_{IR}}{4\omega_{IR}^2} + E_{max}]/\omega_{NIR}.
\end{equation}
Here, $\tilde{I}_{IR}= \alpha I_{IR}$ is an effective peak intensity that we consider by assuming that the single-cycle field did not reach its maximum strength when acting on the propagating free electrons. The parameter $\alpha=$0.22 is a fitting parameter and is chosen to produce the modified cutoff in the presence of the assisting field, and $E_{max}= I_{p} + 3.17I_{NIR}/(4\omega_{NIR}^2)$ is the maximum energy that the ionized electrons can gain in the electric field of the NIR laser. The formula in Eq. (\ref{Nc}), although is simple, it illustrates the origin of the cutoff extension as a result of the excess energy (i.e. 3.17$\frac{\tilde{I}_{IR}}{4\omega_{IR}^2}$) acquired by the electrons from the IR single-cycle field, while are travelling in the NIR laser field. On the other hand, we see the emergence of multiple plateaus, which are found to be extended to higher energies when increasing the intensity of the IR assisting field. These different plateaus are visible in the 1D plot of the HHG spectrum. This is shown in Fig. \ref{fig2}(b) at the intensity of the IR assisting pulse of 1$\times$10$^{13}$ W/cm$^2$. The limit of each plateau is indicated by horizontal dashed lines. The first plateau, which is generated only by the intense NIR pulse is observed for $N_c=$60$^{th}$, as displayed by green color. The presence of the assisting IR pulse results in additional plateaus: the second plateau extends up to the 100$^{th}$ harmonic order and is followed by a third one with an harmonic cutoff at the 160$^{th}$, and even a fourth plateau which is extended up to the 210$^{th}$ but appears with a weak harmonic yield. 

A closer inspection of the spectrum further reveals the emergence of even-order harmonics at the high-order region as a result of including the assisting IR field. This is shown in Fig. \ref{fig2}(c), in which the even harmonics are indicated by vertical lines. In particular it is seen that odd-order harmonics are suppressed in some regions and the harmonic yield is dominated by even harmonics, and in other regions both consecutive odd and even harmonics appear with a comparable yield. We also illustrate the particularity of introducing an IR single-cycle pulse by carrying out a comparison of HHG spectra obtained in the presence of different assisting fields. This is shown in Fig. \ref{fig2}(d), in which the assisting fields are chosen to have the same peak intensity 1$\times$10$^{13}$ W/cm$^2$: XUV-assisting field ($\omega_{XUV}=$41 eV and $\tau_{XUV}=$1 fs), NIR-assisting field ($\omega_{NIR}=$0.976 eV and $\tau_{NIR}=$ 42.4 fs ) and IR single-cycle assisting field ($\omega_{IR}=$0.0976 eV, $\tau_{IR}=$42.4 fs). This comparison shows that the presence of the assisting IR pulse leads to an increase of the harmonic yield by almost 3 order of magnitude compared to other assisting fields. Note that in Figs. \ref{fig2}(c) and (d) the spectra are displayed without using a Gaussian window, which allows a direct comparison of the contribution of the assisting field. On the other hand, the obtained results although are based on a 1D-model, they are found to be well reproduced in a 3D-model as shown in Fig. \ref{fig2}(e), and thus validating the 1D-based calculations.

\begin{figure*}[ht]
\centering
\includegraphics[width=8cm,height=6.cm]{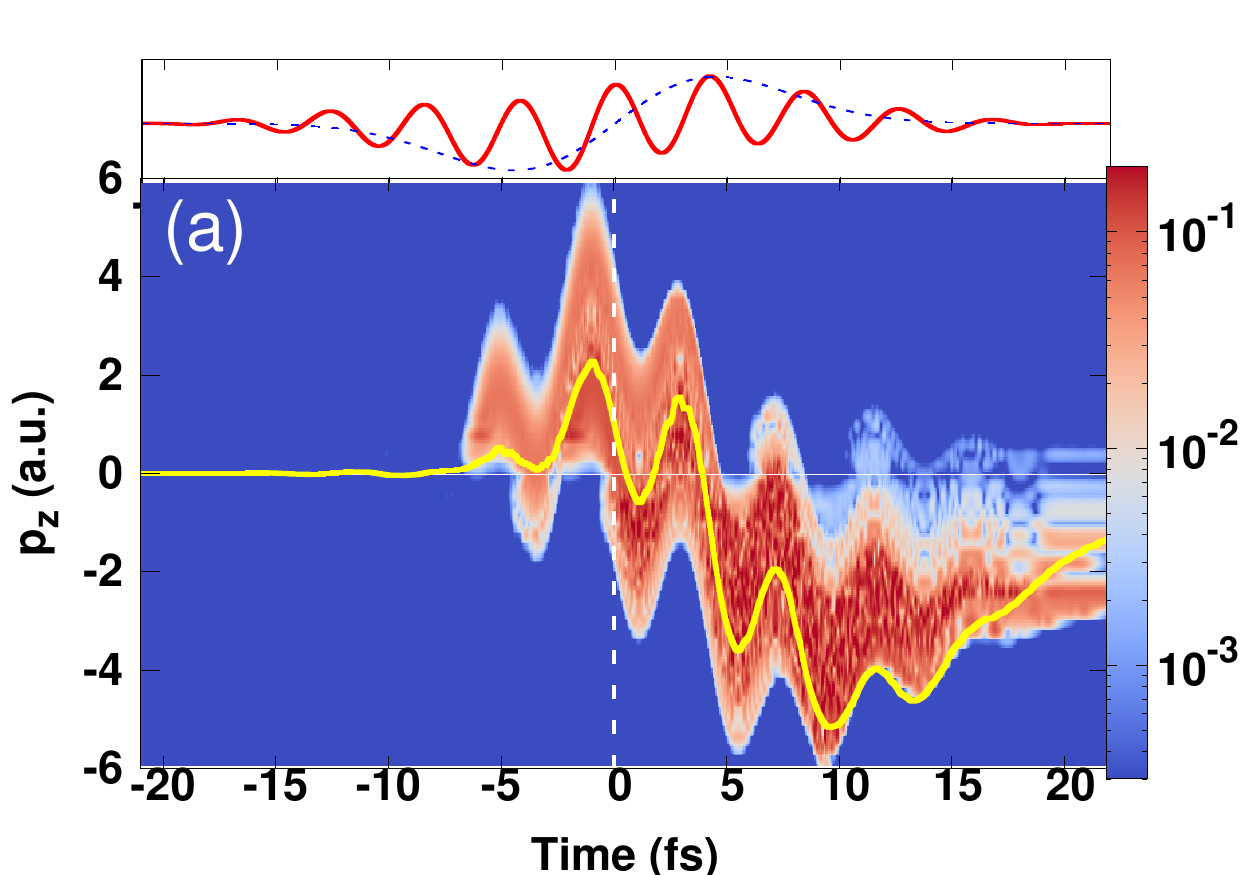}
\includegraphics[width=8cm,height=6.cm]{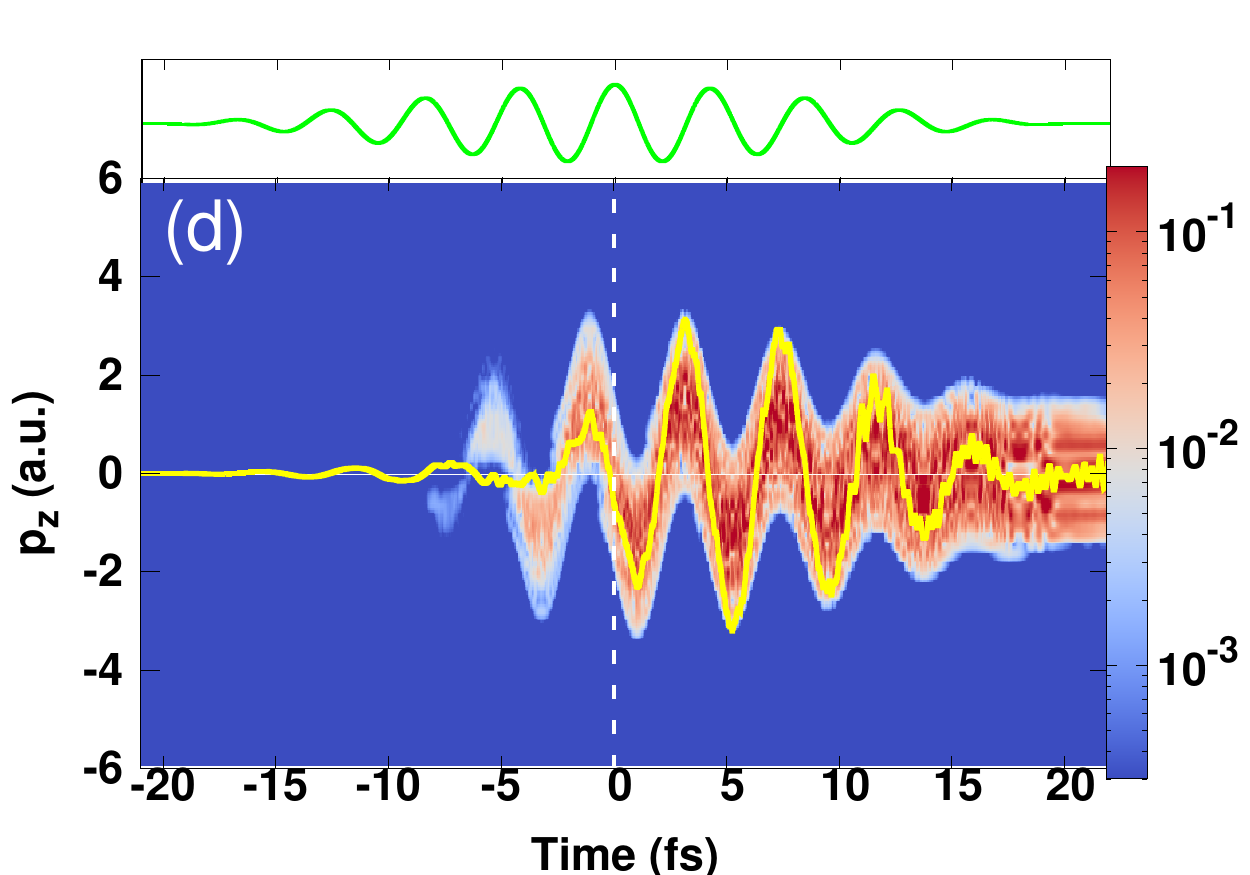}\quad
\includegraphics[width=8cm,height=6.cm]{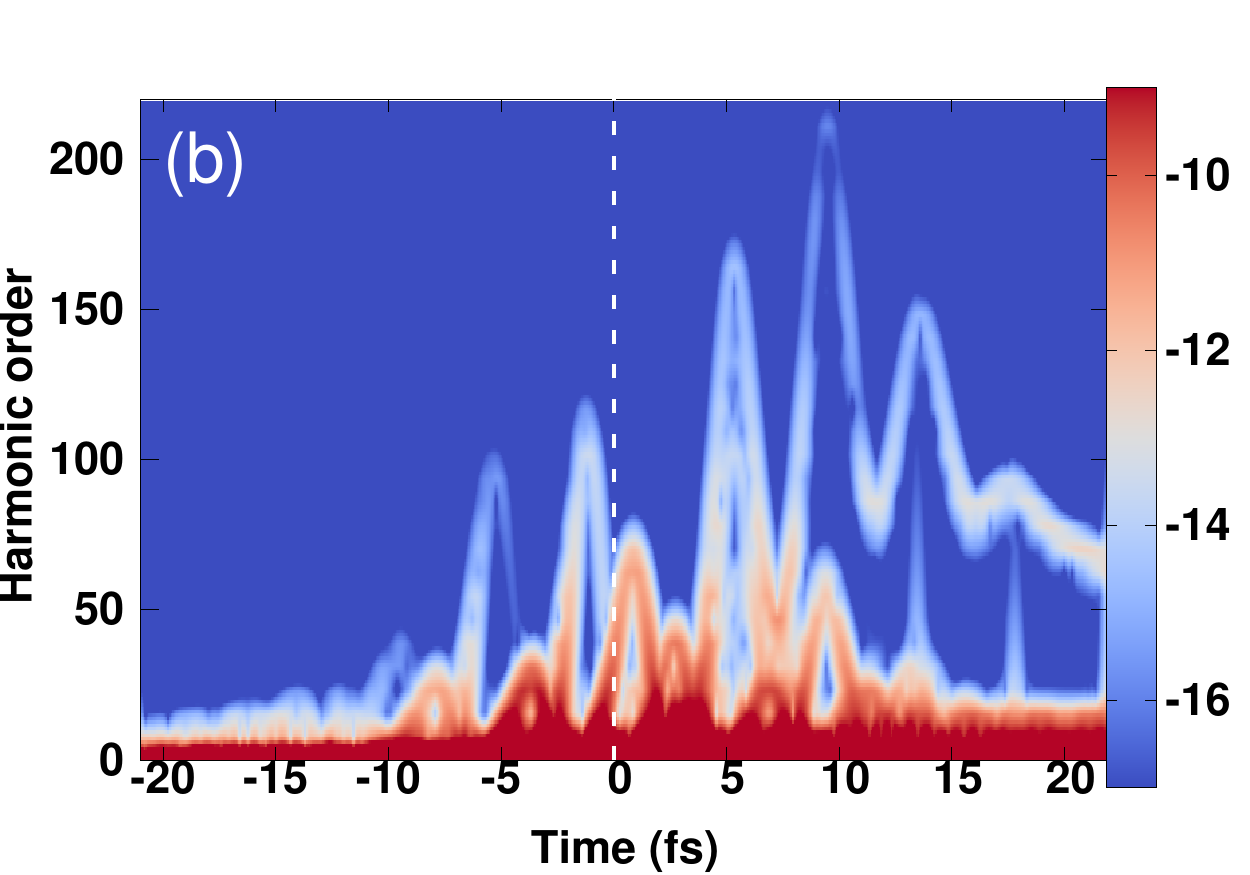}
\includegraphics[width=8cm,height=6.cm]{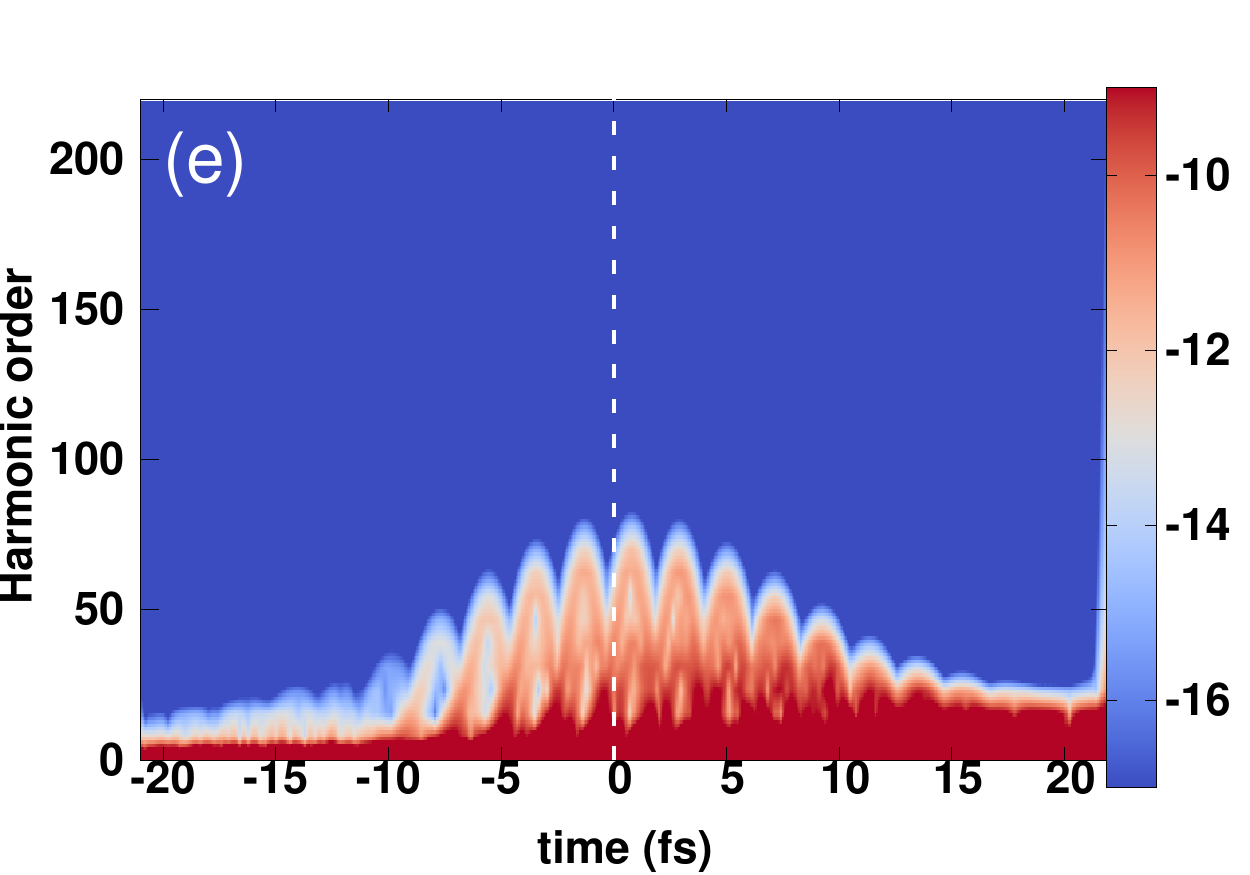}\quad
\includegraphics[width=8cm,height=6.cm]{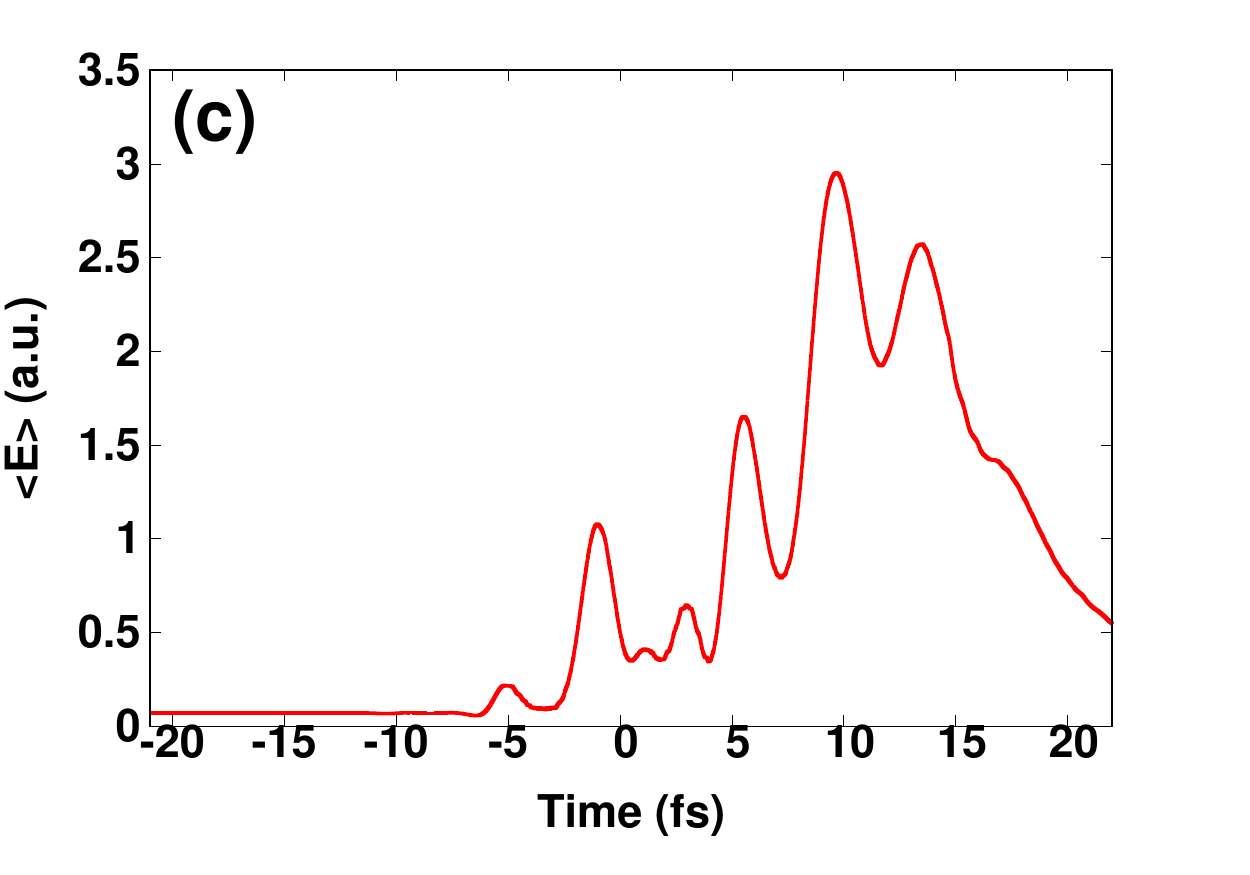}
\includegraphics[width=8cm,height=6.cm]{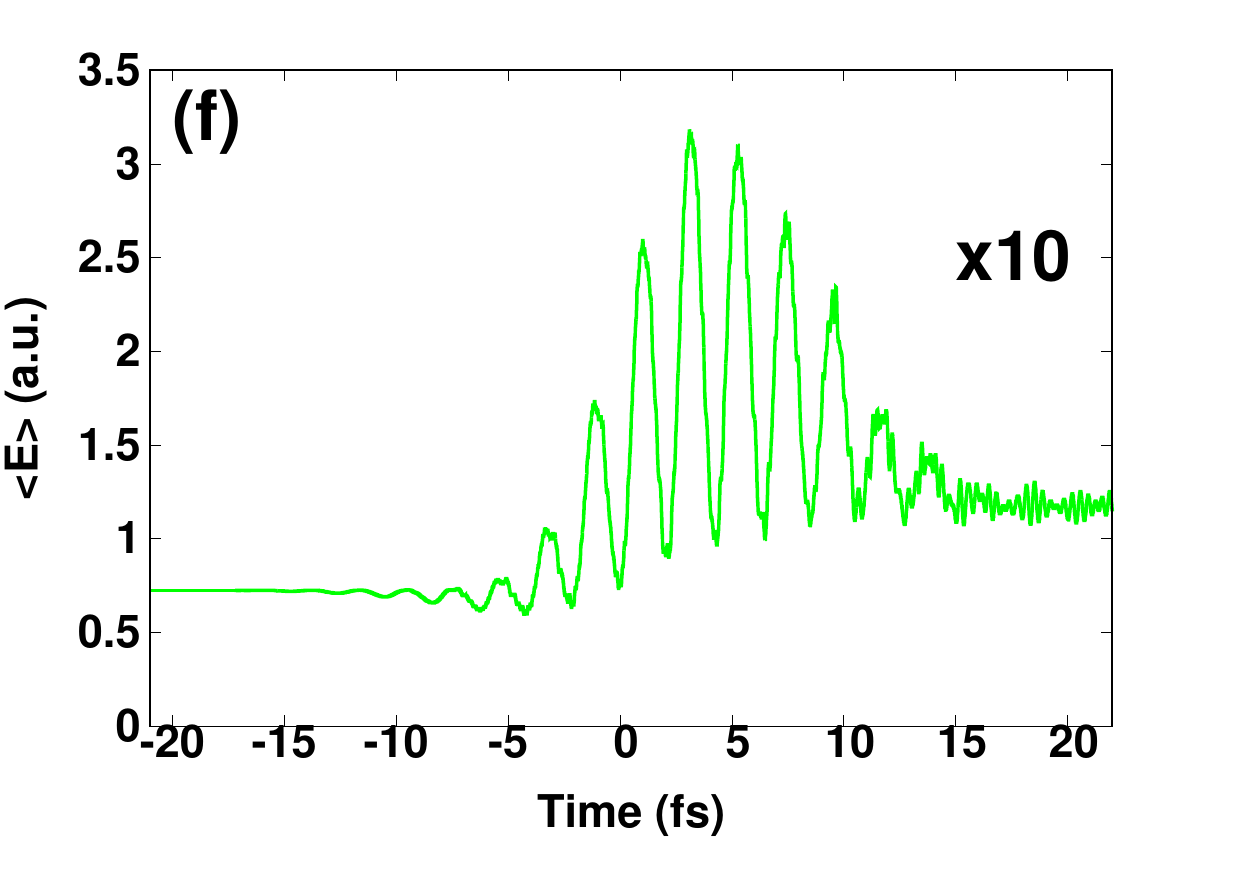}
\caption{\label{fig3} (Color online). 1D-calculations performed with only the NIR pulse (right column) and when assisted with the IR pulse (left column). Top: (a) and (d) Time evolution of the density of the ionized electrons. In the same figures are shown the electron currents (yellow curves). Middle: (b) and (e) Time-frequency analysis of the harmonic spectra. The white dashed lines indicates the zero time. Bottom: (c) and (f) Expectation value of the kinetic energy $<E>$. The data in Fig. (f) are multiplied by 10 due to a weak amplitude of $<E>$. The parameters of the NIR pulse are: $\lambda_{NIR}=$ 1.27 $\mu m$, $T_c=$ 10 cycles, $\delta\phi=$ 0 and $I_{NIR}=$ 1$\times$10$^{14}$ W/cm$^2$. The parameters of the IR pulse are: $\lambda_{IR}=$ 12.7 $\mu m$ and $I_{IR}=$ 1$\times$10$^{13}$ W/cm$^2$. Insets: laser pulses taken from Fig. \ref{fig1}(a).}
\end{figure*}

In order to gain more insights into the physics behind the observed effects, we evaluate the temporal evolution of the density of the ionized electrons presented in momentum space as well as the Gabor time-frequency profile~\cite{Antoine1995}. These are displayed in Fig. \ref{fig3} for both cases: with and without the assisting single-cycle pulse. The results are shown for a peak intensity of the IR pulse of 1$\times$10$^{13}$ W/cm$^2$. At a first glance, the assisting field modifies dramatically the electron density in the forward-backward direction [see Fig. \ref{fig3}(a)], which in turn affects the time-profile of HHG spectra [see Fig. \ref{fig3}(b)]. Indeed, in the first half of the total duration of the pulse (zero-time is indicated by white dashed lines), and in the case the assisting field is introduced, one can see that most of the ionized electrons are distributed in the forward direction, while in the second half the electrons are localized in the backward direction. In both directions, the electrons are produced with high-momenta (up to 6 a.u.). In contrast to the case with the NIR pulse alone [see Fig. \ref{fig3}(d)], the ionized electrons are symmetrically distributed in the forward-backward emission direction, and the maximum momentum produced is around 3 a.u..     

These observations are very important in the sense that they illustrate the role of the IR single-cycle pulse in generating high-momentum electrons and controlling the directionality of the distributed electrons. The origin of the these observations can be understood in the following: During the first half-cycle of the single-cycle pulse, the released electrons by means of the intense NIR pulse receive a high-momentum kick from the single-cycle field, as a result, they get accelerated while travelling in the NIR field and displaced following a unidirectional path, and predominantly end up in the forward direction. In this direction, the electron density follows the strong oscillating NIR field, as can be seen in Fig. \ref{fig3}(a). This is also illustrated in the picture of the electron current, which is here positive, while its sign changes in the case with the NIR pulse alone, as indicated by yellow curves in Figs. \ref{fig3}(a) and \ref{fig3}(d). Here, the generated electrons, while acquiring high-energy from the single-cycle field, they get driven by the strong NIR field to undergo multiple recollisions. The acquired energy is well described by Eq. (\ref{Nc}) and the change of the kinetic energy of the electrons can be seen in the picture of the expectation value $<E>$. The latter helps to measure this change with respect to the reference case, in which the assisting field is absent. The dramatic change of the kinetic energy is found to be almost 9 times higher than that seen in the case with only the NIR pulse, as depicted in Figs. \ref{fig3}(c) and \ref{fig3}(f). Note that the $<E>$ in Fig. \ref{fig3}(f) is multiplied by 10 to make it visible.

\begin{figure*}[ht]
\centering
\includegraphics[width=8cm,height=6.5cm]{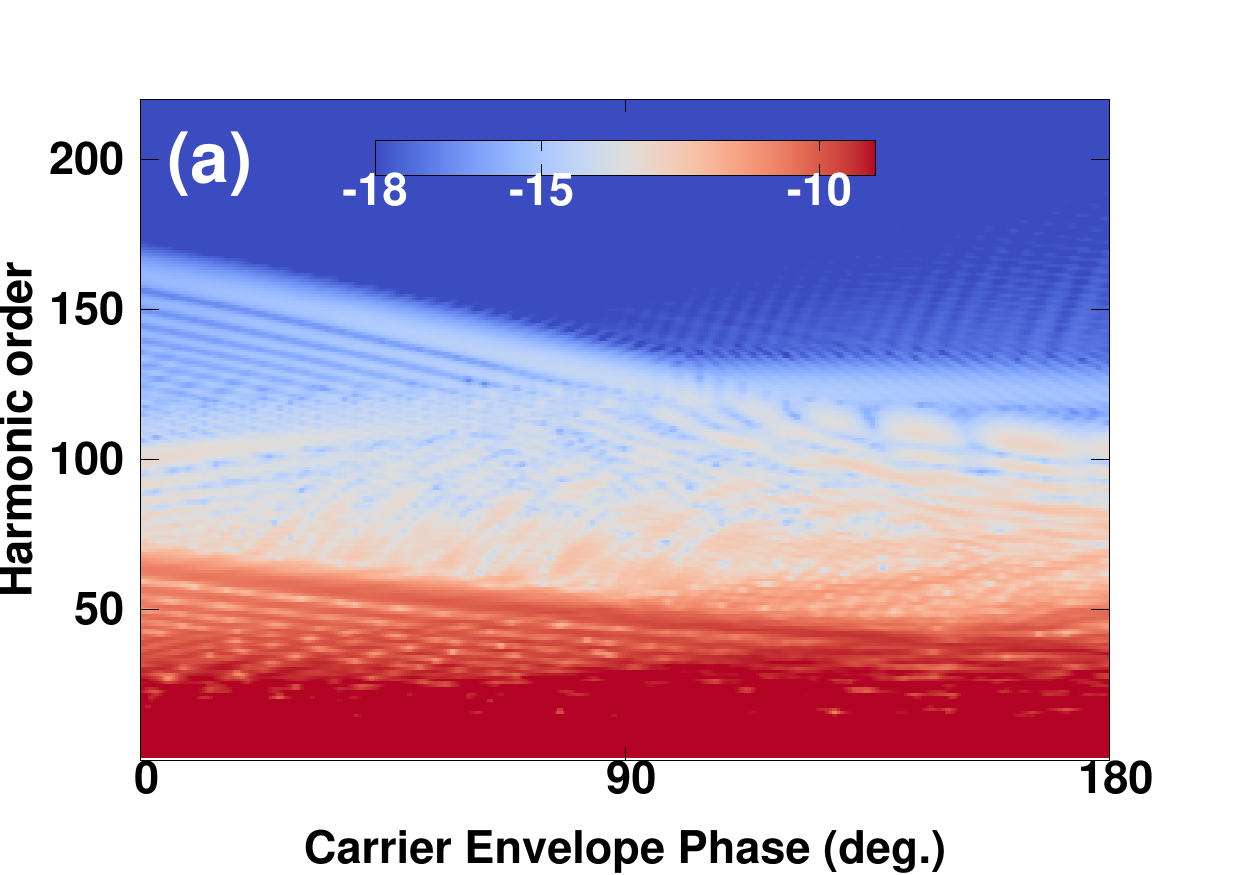}
\includegraphics[width=8cm,height=6.5cm]{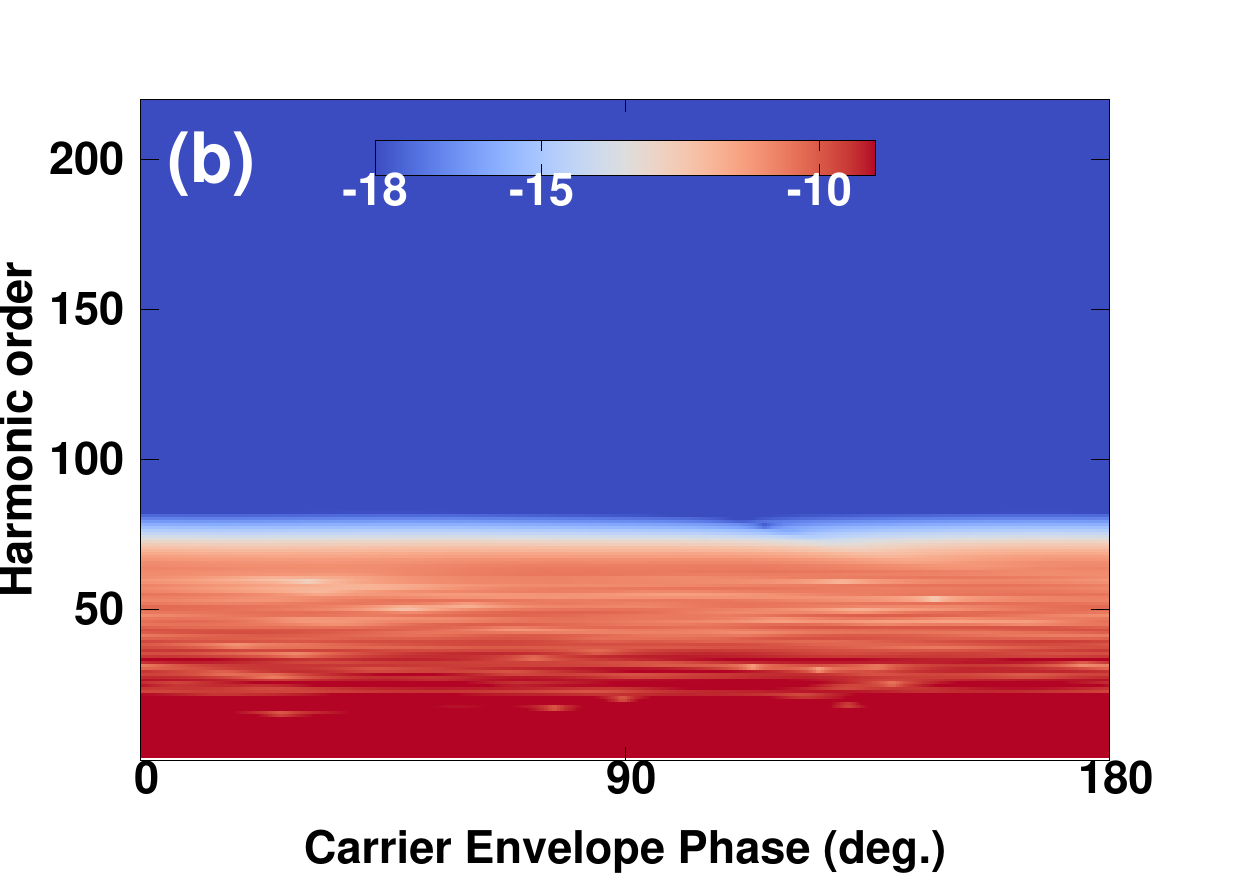}
\caption{\label{fig4} (Color online). 1D-calculations of the harmonic spectra as a function of the relative optical phase between the NIR and IR pulses. (a) In the presence of the assisting IR pulse and (b) in the absence case. The parameters of the NIR pulse are: $\lambda_{NIR}=$ 1.27 $\mu m$, $T_c=$ 10 cycles and $I_{NIR}=$ 1$\times$10$^{14}$ W/cm$^2$. The parameters of the IR pulse are: $\lambda_{IR}=$ 12.7 $\mu m$ and $I_{IR}=$ 1$\times$10$^{13}$ W/cm$^2$.}
\end{figure*}

In the time-frequency profile, the effect of introducing the assisting IR field manifests by the emergence of two emission bursts, in which the harmonic order of the strongest one is located around the 100$^{th}$, and that corresponds to the maximum energy of the second plateau as depicted in Fig. \ref{fig2}(b) with a horizontal dashed line. This result is supported by the calculated expectation value of the kinetic energy $<E>$, which shows two peaks at the same emission times provided by the Gabor profile. On the contrary, this behavior is absent in the case with the NIR pulse alone [see Figs. \ref{fig3}(e) and \ref{fig3}(f)]. Similarly in the second half-cycle: the assisting field reverses its sign and the ionized electrons are displaced in the backward direction and acquire once again high-momenta from the assisting IR field. In this direction, these high-energy electrons are driven by the oscillating NIR field to undergo multiple recollisions. This is reflected in the time-frequency profile [see Fig. \ref{fig3}(b)] by the emission of three bursts, whose harmonic orders are around the 160$^{th}$, 210$^{th}$and 150$^{th}$, respectively in time. These bursts correspond to the emergence of the third and the fourth plateaus, as indicated by horizontal dashed lines in Fig. \ref{fig2}(b). The signal of the fourth plateau is however weak. In contrast, only the basic harmonic components appear in the case of the NIR pulse alone [see Figs. \ref{fig3}(e) and \ref{fig3}(f)]. 

On the basis of these analyses, we conclude that the origin of the emerged high-energy plateaus is caused by the excess energy acquired by the ionized electrons from the IR single-cycle field, which then get driven by the strong oscillating NIR field to undergo high-energy electron recollisions. These findings, therefore, demonstrate the role of a IR single-cycle pulse as an attractive means for producing high-energy electrons and controlling the directionality of the distributed electrons.

Having established a comprehensive picture of the role of a IR single-cycle pulse in inducing high-energy plateaus, we now discuss how this phenomenon can be controlled. Taking advantage of the properties of the optical phase in controlling the continuum wavepacket, as it has been discussed in the context of a single-color few-cycle pulse~\cite{De1998} and within a two-color scheme involving two multi-cycle laser pulses~\cite{Agueny2019,Agueny2020b}), we aim here at implementing this control procedure. The basic physics involved in this control scheme is that the ionized electrons are presumed to follow the instantaneous oscillating field, as shown in Fig. \ref{fig3}(d). In general, changing the optical phase affects the time-birth of the ionized electrons, due to a time-offset expressed as $\tau=\delta\phi/\omega_{NIR}$. Therefore, electrons are generated with different final momenta depending on the optical phase, which in turn affects the maximum energy the electron gain from the laser fields. In the case of a NIR multi-cycle pulse alone, however, the final energies of the electrons are insensitive to the change of the optical phase. This is shown in the picture of the HHG spectrum [see Fig. \ref{fig4}(b)], which is displayed at different optical phases covering the range [0,$\pi$]. Note that the parameters of the two-color scheme are the same as in Fig. \ref{fig3}. When adding the IR single-cycle pulse, the spectrum exhibits a strong sensitivity to optical phase [see Fig. \ref{fig4}(a)]. Here, both the harmonic yield and the cutoff region are found to be modified. In particular, changing the relative optical phase by $\pi/2$ results in an enhancement of the harmonic yield. These modifications are a signature of the ultrafast coherent control of the HHG process. Thus, by varying the relative optical phase one can precisely tailor the optical cycles to yield to an unprecedented degree of control for the characteristic features of the process.

\begin{figure}[h!]
\centering
\includegraphics[width=8cm,height=15cm]{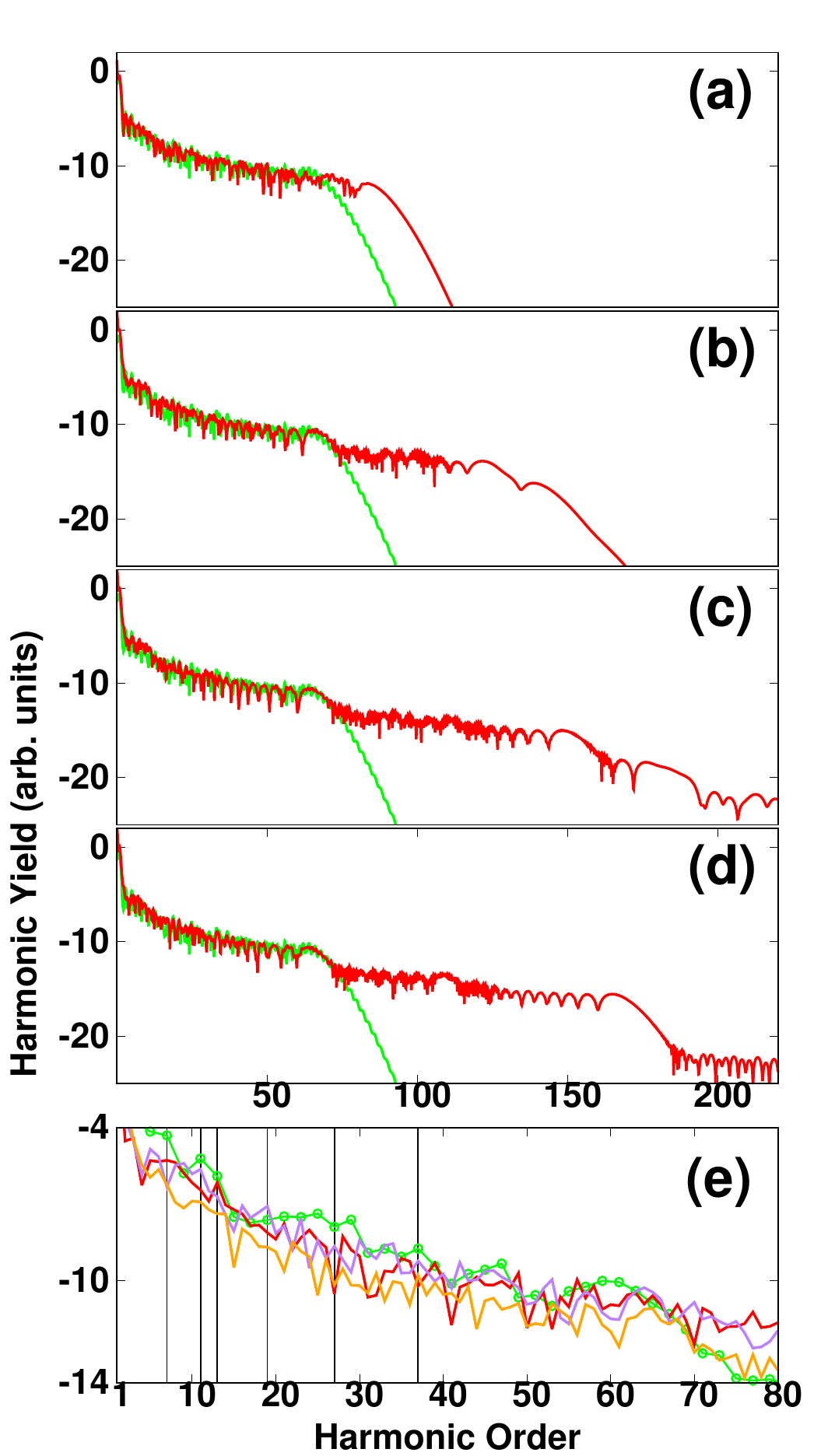}
\caption{\label{fig5} (Color online). 1D-calculations of the harmonic spectra obtained for a peak intensity of the IR single-cycle pulse of 1$\times$10$^{13}$ W/cm$^2$ and at different wavelengths (red curve): (a) $\lambda_{IR}=$ $2\lambda_{NIR}$, (b) $\lambda_{IR}=$ $4\lambda_{NIR}$, (c) $\lambda_{IR}=$ $6\lambda_{NIR}$, and (d) $\lambda_{IR}=$ $8\lambda_{NIR}$. For reference, the HHG spectrum obtained with the NIR pulse alone is shown with green curve in (a)-(d) and also with green empty circles in (e). A zoom of the spectrum in the harmonic order range [1:80] is shown in (e) for: $\lambda_{IR}=$ $10\lambda_{NIR}$ (red curve), $\lambda_{IR}=$ $6\lambda_{NIR}$ (purple) at 1$\times$10$^{13}$ W/cm$^2$. Orange curve shows the spectrum at the IR intensity of 4$\times$10$^{13}$ W/cm$^2$ ($\lambda_{IR}=$ $10\lambda_{NIR}$). Vertical lines indicate the suppressed odd-order harmonics. The parameters of the NIR pulse are: $\lambda_{NIR}=$ 1.27 $\mu m$, $T_c=$ 10 cycles, $\delta\phi=$ 0 and $I_{NIR}=$ 1$\times$10$^{14}$ W/cm$^2$.}
\end{figure}

For completeness, we show in Fig. \ref{fig5} how tuning the wavelength of the single-cycle pulse affects the HHG spectrum. Here, we keep the parameters of the NIR pulse unchanged, and we vary both the wavelength and the peak intensity of the single-cycle pulse covering the spectral range [3,10] $\mu m$. The calculated spectra at the peak intensity of the IR pulse of 10$^{13}$ W/cm$^2$ are depicted with red color and show an extension of the harmonic cutoff when increasing the wavelength, and that is consistent with the formula in Eq. (\ref{Nc}). Here, the harmonic order cutoff is extended from 100$\omega_{NIR}$ in the case of $\lambda_{IR}=$ $2\lambda_{NIR}$ to 160$\omega_{NIR}$ for $\lambda_{IR}=$ $8\lambda_{NIR}$. For reference, the HHG spectrum obtained with the NIR pulse alone is shown with green color. On the other hand, it is seen that the change of the wavelength does not affect the harmonic yield of the first plateau. A closer inspection of the spectrum, however shows some differences that emerge on the location of the harmonic components [cf. Fig. \ref{fig5}(e)]. In particular, the change of the wavelength from $\lambda_{IR}=$ $6\lambda_{NIR}$ (purple curve) to $\lambda_{IR}=$ $10\lambda_{NIR}$ (red curve) leads to a suppression of some odd-order harmonics, for instance the 7th is suppressed in the case of $\lambda_{IR}=$ $6\lambda_{NIR}$ and the 11th is suppressed in the case $\lambda_{IR}=$ $10\lambda_{NIR}$, as indicated by vertical lines in the low-order harmonic region. For reference, odd-order harmonics in the absence of the assisting field is also shown (green curve with empty circles). Insights into these observations are provided by the following expression describing the harmonic yield of a specific order $q$
\begin{equation}\label{Hq}
H_q(\omega_q) = |\int dz |D_z(z,\omega_q)|e^{-i\phi_q(z,\omega_q)}|^2.
\end{equation} 
Here $|D_z(z,\omega_q)|$ is the magnitude of the dipole accelerator [cf. Eq. (\ref{Dzw})] at the point $z$ and $\phi_q$ is the associated phase. According to Eq. (\ref{Hq}) each harmonic $H_q$ results from a coherent sum of different dipole amplitudes $D_z(z,\omega_q)$. These amplitudes might interfere in position space either constructively or destructively depending on the phase $\phi_q$ [cf. Eq. (\ref{Hq})]. The presence of the single-cycle pulse might modify this phase, and that depends on its parameters (i.e the peak intensity and the wavelength). The observations in Fig. \ref{fig5} indicate that the single-cycle pulse affects slightly slow electrons, which are the main source of the generated low order-harmonics, and thus the first plateau. These electrons feel the coulomb field and are governed by the intense NIR multi-cycle pulse. While high-energy electrons, which propagate freely, get accelerated by the single-cycle field. Those electrons are the main source for the build-up of high-energy plateaus. This picture holds as the single-cycle field is not strong enough to undergo ionization. When increasing its peak intensity up to 4$\times$10$^{13}$ W/cm$^2$, one can see that the harmonic yield of the first plateau get modified [cf. Fig. \ref{fig5}(e), orange curve]. In this case the yield diminishes probably due to destructive interference effects between dipole emissions for a specific harmonics $q$. 

These obtained results demonstrate that IR single-cycle pulse affects mainly high-energy electrons. On the other hand, the results indicate that tuning the wavelength of the IR pulse allows one to control the extension of the harmonic cutoff. Therefore, the results presented in this work are valid for a wide spectral range of the IR single-cycle pulse, which demonstrates the generality of its effect on the HHG process.

At this point, our discussion presented in this work elucidates the effect of introducing a weak IR single-cycle pulse to control the HHG process, and particularly to induce high-energy electron recollisions. This is an interesting finding, since it can be exploited for time-resolved electron diffraction in an experiment, and that might lead to establishing an atomic interferometer for imaging in time and space the electron motion. With the state-of-the-art laser technology, it is possible to generate IR single-cycle pulses with peak intensities and in the spectral range discussed here (see e.g. references~\cite{Nie2018,Zhu2020}), in which relativistic infrared single cycle pulses have been produced.

\section{CONCLUSIONS}\label{conclusions}

To conclude, on the basis of numerical simulations of the 1D-TDSE, which we verify using a 3D-model, we show that the HHG process induced by an intense near-infrared multi-cycle pulse can be controlled efficiently using an infrared single-cycle pulse. This control scheme manifests in the HHG spectrum by the generation of even-order harmonics and by an extension of the harmonic cutoff by a factor of 3 accompanied with the appearance of high-energy plateaus. The origin of this emerged plateaus is found to be related to a vast momentum kick that ionized electrons receive from the single-cycle field. These electrons in turn get accelerated to high-momenta and displaced following a unidirectional path, thus leading to high-energy electron recollisions. We also identify the role of the single-cycle field in controlling the directionality of the released electrons, which are found to be displaced mainly in the backward direction at the end of the pulses. Furthermore, it is found that these emerged effects exhibit a strong sensitivity to the change of the relative optical phase between the two pulses as well as the wavelength of the single-cycle pulse. This sensitivity is shown to be a signature of ultrafast coherent control of the HHG process. Thus, our findings establish a new control scheme, so far unexplored in strong-field physics, which might open up a new route for time-resolved electron diffraction by means of an infrared single-cycle field-assisted HHG.

\section*{ACKNOWLEDGMENTS}
The computations based on the 3D-model were performed on resources provided by UNINETT Sigma2 - the National Infrastructure for High Performance Computing and Data Storage in Norway. The author A.T acknowledges support from the bilateral relationships between Morocco and Hungary in Science and Technology (S \&T) under the project number 2018-2.1.10-TET-MC-2018-00008.



%
\end{document}